\documentclass{elsart}
\usepackage{epsfig}

\usepackage{amssymb}
\newcommand{\Gammastar}{\Gamma^* }
\newcommand{\xibar}{\overline{\xi} }

\begin{document}

\begin{frontmatter}



\title{Clustering in N-Body gravitating systems}
\author[Uni,infm]{Maurizio Bottaccio\corauthref{cor1}}
\ead{mb@pil.phys.uniroma1.it}
\author[Uni,infm]{Luciano Pietronero}
\author[Uni]{Alessandro Amici}
\author[Uni]{Paolo Miocchi}
\author[Uni]{Roberto Capuzzo Dolcetta}
\author[infm]{Marco Montuori}
\corauth[cor1]{Corresponding author. Tel.:+39 0649913450; fax: +39 064463158}

\address[Uni]{Dep. of Physics - Univ. di Roma ``La Sapienza'' -  
P.le A. Moro, 2 -  00185  Rome (Italy)}

\address[infm]{INFM -  Research Unit  Roma 1 -  
  Rome (Italy)}

\begin{abstract}
Self-gravitating systems have acquired growing  
interest in statistical mechanics, due to the peculiarities
 of the 1/r potential. Indeed, the usual approach of statistical
 mechanics cannot be applied to a system of many point particles
 interacting with the Newtonian potential, because of (i) the long 
range nature of the 1/r potential and of (ii) the divergence at
 the origin. We study numerically the evolutionary behavior of 
self-gravitating systems with periodical boundary conditions, 
starting from simple initial conditions. We do not consider
in the simulations additional effects as the (cosmological) metric
 expansion and/or sophisticated initial conditions, since we are
 interested whether and how gravity by itself can produce
 clustered structures. We are able to identify well defined
 correlation properties during the evolution of the system,
 which seem to show a well defined thermodynamic limit, as
 opposed to the properties of the ``equilibrium state''.
 Gravity-induced clustering also shows interesting self-similar
 characteristics. 
\end{abstract}

\end{frontmatter}

\section{Introduction}

Many-body systems with long range interactions are widespread in nature.
In particular,
 gravitating systems dominate in  an impressive range of spatial
scales ($10^{-1}pc - 10^8pc$) .
It is very difficult to trace the specific role of gravity
in the evolution and the statistical properties of such systems,
since usually many different physical processes should be considered.
For this reason, many theoretical and numerical studies 
have focussed on the specific physical problem, rather than on 
the general statistical features of an infinite gravitating system.
On the other hand, the usual statistical and thermodynamical approaches
to the understanding   of an infinite system with an interaction of 
some kind fail, or have to be taken very carefully, in the case of
long range interactions.
One of the properties of such interactions is that they make  the system 
{\em non extensive}, that is, the system cannot be considered as
made of smaller independent parts. Therefore one cannot study the
typical properties of 
finite sample to extrapolate safely
 the properties of the whole infinite sample, as one does with
short range systems.
Many simple idealized examples have been considered to analyse
the effect of non extensivity (e.g. \cite{torcini,ruffo,latora,thirring}). 
They have given interesting 
clues on the behavior of non extensive system. On the other hand,
it would be more satisfying to deal with three dimensional
systems with a real interaction potential, as gravity,
 and a full treatment of the dynamics of the system.
Our approach has been to deal with a system as realistic as possible,
i.e. three dimensional N-body gravitating systems


\section{The simulations}

The model we investigate is usually a system
with particles placed randomly in a cube with
periodical boundary conditions. The particles 
have the same mass and are initially at rest.
Since we want to understand the behaviour 
of an infinite system, we run a series of simulations
with different number of particles, but with the same 
density.
It is important to stress that, due to 
long range nature of gravitational interaction,
one has to use subtle analytic tools to evaluate
the force on a particle {\em due to an infinite system}.
We use Ewald formula for this pourpose \cite{ewald}.
It is also important to observe that the gravitational
potential in such systems is not well defined, since
the potential energy of a particle would diverge.
Neverthless, due to isotropy of the system  on
large scales, the force {\em is} well defined,
since contributions from far away are balanced.
We use a numerical code \cite{miocchi} which solves the
$N$ coupled equations of motions in discrete time
steps. A well known algorithm ({\em tree code}) is used
to evaluate the forces in an efficient approximated way.
The time integrator is a {\em leapfrog} (Verlet) algorithm,
with individual time steps.
For numerical reasons, the potential is smoothed at very small scales
(much smaller than the mean interparticle distance),
that is the divergence in the origin is removed and replaced with
a smoother behavior. 
We have checked by changing the smoothing length that such procedure
does not affect our results.

\begin{figure}
\centerline{
        \psfig{file=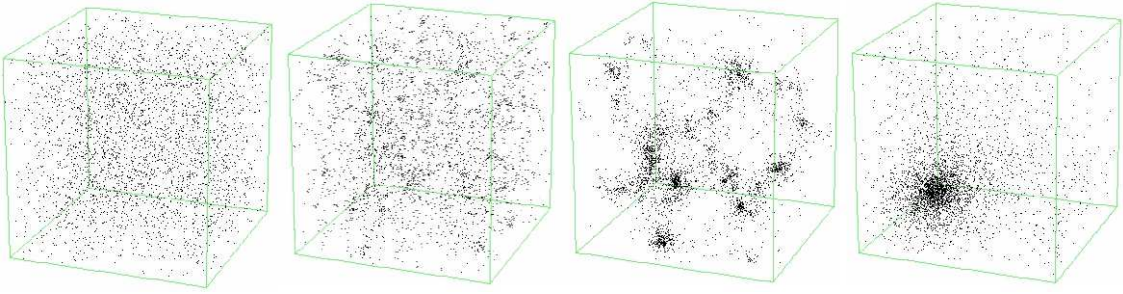,width=15cm}
}
\caption{Evolution of particle positions for a simulation with $32^3$
particles. Time flows from left to right.
}
\label{fig:evol}
\end{figure}

We consider in the following the results of 
a series of simulations, with the same number density
but different number of particles, ($8^3$, $16^{3}$, $32^{3}$).
In fig.\ref{fig:evol} we show a picture of the evolution of 
particle positions in time. Note that the systems develop
clusters on small scale first, then on larger and larger
scales. When the typical size of the clusters is of the order of the box
size, the evolution is influenced by finite size effect. Actually,
even if the periodic system is infinite, the maximum number of particles
in the simulation volume is fixed.

\begin{figure}
\centerline{
        \psfig{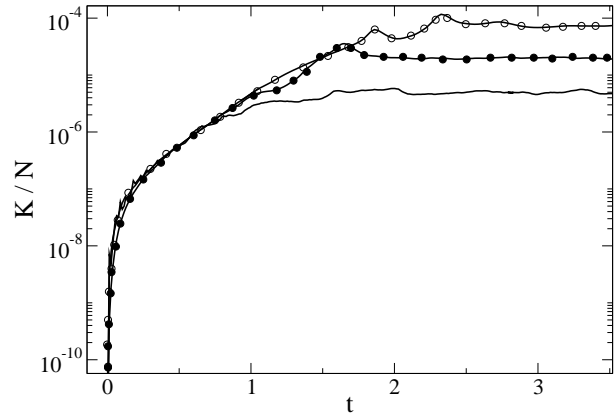}
}
\caption{Evolution of the kinetic energy per particle for simulations
with $8^3$, $16^3$ (filled circles), $32^3$ (empty circles) particles.
}
\label{fig:kin_evol}
\end{figure}

\begin{figure}
\centerline{
        \psfig{file=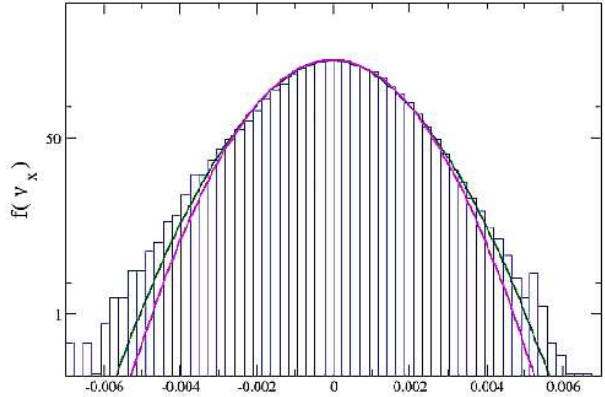,width=8cm}
}
\caption{Normalized distribution of velocities along an axis for a $32^3$ particle simulation. The inner continuous line is the gaussian fit, the outer
one is a stretched exponential fit.
}
\label{fig:vel_distr}
\end{figure}

\begin{figure}
\centerline{
        \psfig{file=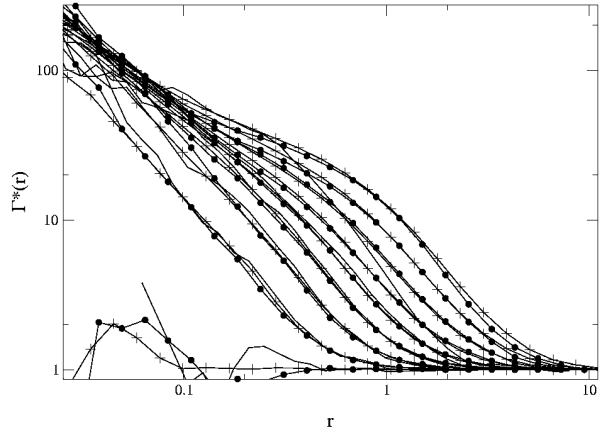,width=8cm}
}
\caption{ Evolution of $\Gammastar (r,t)$ for simulations 
with $8^3$, $16^3$ (lines with pluses), $32^3$(lines with filled circles)
 particles.
Time flows from left to right.
}
\label{fig:gamma}

\end{figure}

In fig.\ref{fig:kin_evol} we plot the kinetic energy per particle versus
time. This figure
 shows that the evolution
of the kinetic 
energy per particle is {\em independent of the number of particles} in the
system, in spite of the non extensivity, up to a time which 
increases approximately with $ln N$. After this time, the kinetic energy
becomes constant. This ``equilibrium state'', though, is a finite size effect.
Therefore we
argue that in the ``thermodynamic limit'' (defined as the limit behavior
when $N , V\rightarrow\infty$ keeping the density constant) the 
behavior of the kinetic energy will be the same. 
It is also interesting to look at the  distribution  of velocities
 along an axis at a fixed time, shown in fig.\ref{fig:vel_distr}.
The picture refers
to a given time before the finite size effects become relevant.
A gaussian fit is shown for the reference.
The distribution is  {\em non gaussian}, but this is not surprising,
 since the system
is not in equilibrium. It can be interesting to try to apply
Tsallis' temptative velocity distribution  (see e.g. \cite{tsa}) on such data.
We have also analysed clustering by means of the correlation functions.
In particular we use the integrated conditional density
$\Gammastar(r,t)$ \cite{pietro} and
the integrated two point correlation function $\xibar (r,t)$ .
Operatively they can be defined respectively as the mean average density
and the mean average density fluctuation in a sphere of radius $r$ 
{\em centered on a particle of the system}.
Therefore a system is clustered on a scale $r_0$ if $\Gammastar (r_0,t)>> n_0$,
where $n_0$ is the mean number density.
The evolution of $\Gammastar (r,t)$ 
is shown in
 fig.\ref{fig:gamma} 
for simulations with different number of particles. Two important conclusions
can be drawn from fig.\ref{fig:gamma}. First of all, the growth of correlations
is approximately the same for all the simulations,
 in the regime where clustering takes
places on scales small compared to the box size. Together with fig.\ref{fig:kin_evol},
this result suggests that the {\em dynamics} in this regime is the same
for any number of particles, and therefore also in the ``thermodynamic limit''.
We can also reasonably infer that the infinite system
has {\em no equilibrium state}.
Another important remark is that the functional form of the $\Gammastar (r,t)$
on the largest scales on which clustering is taking place is {\em the same} 
at all times. So we suppose that the clustering process could be
explained in terms of the same dynamical processes, acting as time goes on on
 larger and larger scales.

In two papers of ours \cite{mau1,mau2}
 we have tried to explain the whole evolution
of the correlations in such systems in terms of a discrete dynamics
acting on larger and larger scales. 
At the beginning, the basic elements for the dynamics we consider are
particles, which form clusters. At later times, the interactions between 
such clusters are mainly responsable
for the further evolution of the system, and so on.
This is a sort of {\em renormalization} of the dynamics.
The justification behind such model lies on the fact that
on scales much larger than the clustering scales the system is approximately
isotropic, therefore the force due to far away objects almost balances.
On the contrary, on the scale of clustering the system is {\em anisotropic},
therefore the force due to near clusters is  large. In particular, it
will be mostly in the direction of the nearest cluster.
\begin{figure}
\centerline{
        \psfig{file=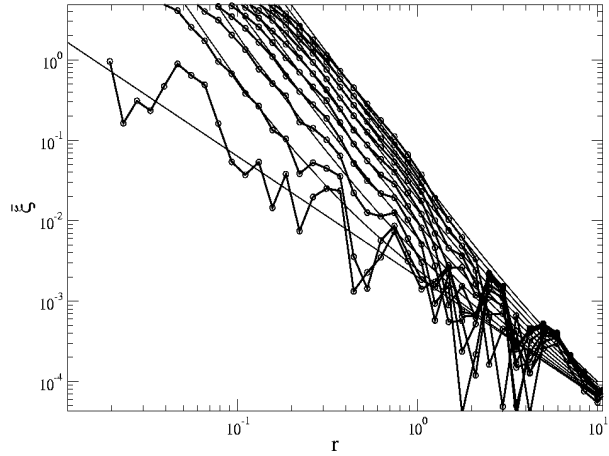,width=8cm,height=6cm}
}
\caption{Comparison between $\xibar$ evolution for a simulation with
$32^3$ particles (filled circles lines) and our theoretical prediction 
(smooth lines). 
The agreement for times
smaller than the typical crossing time is quite good.
}
\label{fig:xi-fit}
\end{figure}
This model is successful in providing:
( i ) an analytical expression for early time 
evolution of correlations, with no free parameter. The agreement with 
numerical experiments is given in
fig.\ref{fig:xi-fit};
( ii ) a qualitative prediction for the rate of the 
clustering process
 at any time.

\section{Conclusions}
We analyse the problem of the evolution of an infinite system with
long range (gravitational) interaction.
We have studied it by numerical N-body simulations of periodical systems
with different sizes. 
We identified some properties which  can be inferred for the infinite system
and we have given a model that could explain most of them.
The relevant points of our approach are:
\begin{itemize}
\item We consider a thermodynamic approach to be highly problematic in the
case of infinite self gravitating systems, since they presumably have no
equilibrium state and they are non-extensive
\item We propose  a
fully dynamical model instead which explains
 the out-of-equilibrium evolution of the
system.
The fundamental elements for the dynamics in our approach
 are discrete objects.
Discretisation turns out to be crucial to explain highly developed clustering,
while standard (statistical mechanical or cosmological) approaches have mostly
focussed on mean field or fluid like  dynamics.In particular, the 
non-analiticity of the distribution of mass (that is, its discrete nature)
is exported at larger and larger scale.
\item The evolution of clustering on a scale depends on what has happened
on smaller scales. This is an opposite view with respect to the
hierarchical clustering model, often considered in astrophysics.
In particular, in such model clustering on a scale depends on initial
 conditions on that scale.
\end{itemize}

\end{document}